\documentclass[11pt, nonacm]{acmart}

\usepackage[most]{tcolorbox}
\usepackage{caption}
\usepackage{array}
\usepackage{diagbox}
\usepackage{hyperref}
\usepackage{supertabular}
\usepackage{booktabs}
\usepackage{float}
\usepackage{enumitem}
\usepackage{graphicx}
\usepackage{comment}
\usepackage{multirow}
\usepackage{tabularx}
\usepackage{colortbl}
\usepackage{longtable}
\usepackage[resetlabels,labeled]{multibib}
\usepackage{subcaption}
\usepackage{makecell}
\usepackage{lscape}

\acmJournal{CSUR}

\setcopyright{none}
\renewcommand\footnotetextcopyrightpermission[1]{}

\setcopyright{none}
\renewcommand\footnotetextcopyrightpermission[1]{} 

\AtBeginDocument{%
  \providecommand\BibTeX{{%
    \normalfont B\kern-0.5em{\scshape i\kern-0.25em b}\kern-0.8em\TeX}}}

\begin{document}

\title{A Survey on Immersive Cyber Situational Awareness Systems}

\author{Hussain Ahmad}
\email{hussain.ahmad@adelaide.edu.au}

\author{Faheem Ullah}
\email{faheem.ullah@adelaide.edu.au}

\affiliation{
 \institution{School of Computer and Mathematical Sciences, The University of Adelaide}
 \country{Australia}
}


\author{Rehan Jafri}
\email{Rehan.Jafri@Honeywell.com}

\affiliation{
 \institution{Honeywell}
 \country{United Kingdom}
}

\authorsaddresses{
Authors' addresses: Hussain Ahmad, hussain.ahmad@adelaide.edu.au; Faheem Ullah, faheem.ullah@adelaide.edu.au; Rehan Jafri, Rehan.Jafri@Honeywell.com}

\renewcommand{\shortauthors}{Ahmad et al.}

\begin{abstract}

Cyber situational awareness systems are increasingly used for creating cyber common operating pictures for cybersecurity analysis and education. However, these systems face data occlusion and convolution issues due to the burgeoning complexity, dimensionality, and heterogeneity of cybersecurity data, which damages cyber Situational Awareness (SA) of end-users. Moreover, conventional ways of human-computer interactions, such as mouse and keyboard, increase the mental effort and cognitive load of cybersecurity practitioners, when analyzing cyber situations of large-scale infrastructures. Therefore, immersive technologies, such as virtual reality, augmented reality, and mixed reality, are employed in the cybersecurity realm to create intuitive, engaging, and interactive cyber common operating pictures. The Immersive Cyber Situational Awareness (ICSA) systems provide several unique visualization techniques and interaction features for the perception, comprehension, and projection of cyber SA. However, there has been no attempt to comprehensively investigate and classify the existing state of the art in the use of immersive technologies for cyber SA. Therefore, in this paper, we have gathered, analyzed, and synthesized the existing body of knowledge on ICSA systems. In particular, our survey has identified visualization and interaction techniques, evaluation mechanisms, and different levels of cyber SA (i.e., perception, comprehension, and projection) for ICSA systems. Consequently, our survey has enabled us to propose: (i) a reference framework for designing and analyzing ICSA systems by mapping immersive visualization and interaction techniques to the different levels of ICSA; (ii) future research directions for advancing the state-of-the-art on ICSA systems; and (iii) an in-depth analysis of the industrial implications of ICSA systems to enhance cybersecurity operations.

\end{abstract}

\ccsdesc[500]{General and reference~Surveys and overviews}
\ccsdesc[500]{Security and privacy~Software and application security}
\ccsdesc[500]{Human-centered computing~Mixed / augmented reality}

\keywords{Virtual Reality, Augmented Reality, Mixed Reality, Extended Reality, Cybersecurity, Cyber Situational Awareness, Cybersecurity Education, Cybersecurity Training}
\maketitle

\section{Introduction} \label{introduction}

\newcolumntype{C}[1]{>{\centering\arraybackslash}m{#1}}

With the drastic increase in cybercrime, especially, after the outbreak of the COVID-19 pandemic \cite{williams2020cybersecurity}, cybersecurity awareness, education, and training become imperative for every individual. The COVID-19 pandemic increased the cybercrime rate by 400\% \cite{arogbodo2022impacts}. A record-breaking number of data breaches (i.e., 1862 incidents) are reported in 2021 \cite{DataBreach}. The cost of cybercrime is estimated to be around \$9.22 trillion in 2024 and is expected to reach \$17.9 trillion by 2030, almost doubling over this period \cite{hayetcybersecurity}. Due to the burgeoning cyber incidents, it is envisaged that the global information security market will reach \$376.32 billion in 2029 \cite{forecast}. Moreover, the emergence of new software architectures and digital technologies has introduced a new set of cybersecurity vulnerabilities, reshaping the cybersecurity landscape \cite{jayalath2024microservice, abdulsatar2024towards, ullah2019architecture, ahmad2023review, ahmad2024smart, ahmad4918202towards}. Furthermore, humans play an integral role in cybersecurity \cite{evans2016human}. Many cybersecurity incidents are caused by the lack of cybersecurity education and training of end-users \cite{alqahtani2020design}. For example, the \textit{World Economic Forum} claims that 95\% cyber incidents are caused by human errors \cite{worldEconomicForum}. In addition, as reported by \textit{Cisco}, 42\% of cybersecurity practitioners suffer from cyber fatigue that causes negligence in cybersecurity operations \cite{Fatigue}. Therefore, the area of cyber Situational Awareness (SA) is getting immense attention from learners, analysts, and experts in the cybersecurity realm.

Cyber SA can be considered as an application of Endsley's SA reference model \cite{endsley1988design} in the cybersecurity domain. Cyber SA refers to the identification, collection, analysis, and evaluation of cybersecurity data from a given system to make effective decisions for responding to potential cyber threats \cite{barford2010cyber}. Traditional cyber SA systems provide perception, comprehension, and projection of cyber environments to end users through two-dimensional displays (e.g., 2D screens) with limited interaction capabilities (e.g., mouse, keyboard, and monitor). These conventional visualization and interaction technologies limit users' cognition and understanding of cyber situations, which eventually deteriorates cyber SA. For example, cybersecurity data complexities (e.g., fast, dynamic, and unpredictable data) and heterogeneity lead to data occlusion and convolution in traditional cybersecurity visualizations \cite{sukhija2019employing}, which limits the perception of cyber situations \cite{mattina2017marcs}. Moreover, users suffer from mental effort and cognitive load when they need to shift their focus between multiple terminal windows to understand cyber situations and perform parallel activities through traditional visualization and interaction techniques of cyber SA systems \cite{beitzel2016cognitive}. This issue was spotlighted during the surge of the COVID-19 pandemic when operators needed to monitor pandemic-related data and respond to cybersecurity alerts simultaneously \cite{korkiakoski2021using}.

To solve the abovementioned problems, immersive technologies based on extended reality are used to create effective cyber SA for end-users \cite{munsinger2023virtual, alnajim2023exploring, abu2024enhancing}. These technologies are revolutionizing various domains: in healthcare, they facilitate advanced surgical simulations and interactive therapies \cite{qu2022review}; in education, they augment learning through engaging virtual experiences \cite{sandoval2024systematic}; and in retail, they provide virtual try-ons, significantly improving customer satisfaction and reducing return rates \cite{erensoy2024consumer}. Similarly, in the cybersecurity realm, Immersive Cyber Situational Awareness (ICSA) systems refer to software and hardware systems that allow users to replace or expand their physical environments with virtual objects to create perception, comprehension, and projection of cybersecurity data for a given system. ICSA systems provide several features for enhancing the cyber SA of end-users \cite{skorenkyy2021use}. For example, ICSA systems present multi-dimensional cybersecurity data through different visualization techniques (e.g., metaphorical shapes, icons, and scatterplots) with an ability to arrange its multiple views through natural interaction features (e.g., gaze, gesture, and controller) in a spatial three-dimensional space around users for monitoring, analyzing and forecasting cyber situations. Moreover, an immersive single 3D display of cybersecurity information provides a holistic cyber common operating picture that eradicates distractions and the need for traditional multiple views of cybersecurity data for getting cyber SA. Furthermore, ICSA systems provide engagement, entertainment, and enjoyment in cybersecurity education and training processes \cite{puttawong2017vrfiwall}, which significantly increases cyber SA with less cognitive load and mental effort. With adapting \textit{Industrial Revolution 4.0}, ICSA systems can be employed in a variety of settings including Security Operation Centers (SOCs), operations of Computer Emergency Response Teams, Incident Response Management, Network Security Operations, Computer Network Defense, as well as in cybersecurity education and training. 

Given the advantages of ICSA systems over traditional cyber SA systems, researchers and practitioners have been putting a lot of effort into analyzing, designing, and evaluating visualization and interaction features of ICSA systems for a better understanding of cyber SA. Therefore, the body of knowledge on ICSA systems has continuously been scatteredly expanding. Hence, we decided to collect, analyze, and synthesize the existing literature for systematizing and classifying the state of the art on ICSA systems. In this study, we have extracted, analyzed, and reported the existing literature on ICSA systems. In particular, our survey aims to review the visualization and interaction techniques, evaluation mechanisms, and levels of cyber SA achieved for ICSA systems. The ICSA visualization and interaction techniques are then further analyzed in terms of the perception, comprehension, and projection phases of cyber SA to propose a framework for designing and analyzing ICSA systems. Furthermore, this survey identifies future research directions for researchers and provides industrial insights for practitioners regarding ICSA systems. It is important to mention that we considered immersive technologies based on extended reality, including virtual reality, augmented reality, and mixed reality, in the context of ICSA systems. Other related concepts, such as immersive simulations, gamification, digital twins/shadows, three-dimensional displays, and metaverse, are out of the scope of this survey.
 
\noindent \underline{\textbf{Our Contributions:}} In summary, our survey makes the following contributions.

\begin{itemize}[leftmargin=*]

    \item It provides an overarching analysis of ICSA visualization and interaction techniques identified in the literature. Each visualization and interaction technique is reported in the context of perception, comprehension, and projection of ICSA systems. The visualization and interaction techniques are categorized based on a novel taxonomy separately.
    
    \item It gives a comprehensive analysis of the SA levels achieved for ICSA in the literature. For the first time, each level of SA is described in the context of ICSA. The existing literature is categorized based on the defined levels of ICSA (i.e., perception, comprehension, projection).
   
    \item It presents a high-level investigation of evaluation mechanisms used to validate ICSA systems. Each evaluation mechanism has been critically examined focusing on ICSA usability evaluation, user demographics, performance, and cognition metrics. The evaluation mechanisms are categorized based on the methodology employed for validating ICSA systems.  

    \item It presents a combined analysis of ICSA visualization/interaction techniques, evaluation mechanisms, and levels of ICSA. This thorough analysis leads to the development of a reference framework for designing and evaluating ICSA systems. Additionally, the analysis suggests future research directions and highlights the industrial implications of ICSA systems.
    
\end{itemize}

\underline{\textbf{Survey Structure: }} The rest of this survey is organized as follows. Section \ref{Methodology} presents the research methodology to conduct this survey study. Section \ref{visualization&Interactions}, Section \ref{ICSA_Levels}, and Section \ref{evaluation} report ICSA visualization/interaction techniques, evaluation mechanisms, and levels of ICSA, respectively. Section \ref{discussion} proposes the developed framework based on the analysis of our research findings. This section also describes the potential future research directions and industrial implications for ICSA systems. Lastly, the conclusion of this survey is presented in Section \ref{conclusion}.

\section{Research Methodology}  \label{Methodology}

This section reports the research methodology used to conduct this survey study. We extract, analyze, and synthesize the existing state of the art on ICSA systems using the following five-step research methodology.

\subsection{Research Questions}

This survey aims to provide an overview of the existing state-of-the-art on ICSA systems. We designed a set of three Research Questions (RQs) to review the existing literature on ICSA systems. \textcolor{blue} {Table~\ref{tab:questions}} presents the research questions along with their motivations.

\renewcommand{\arraystretch}{1}
\begin{table*}[b]
\captionof{table}{Research questions of this survey.} 
\label{tab:questions} 
\begin{center}
    {\fontsize{9}{11}\selectfont
    \begin{tabular}{|m {22em}|m{22em}|}
         \hline
         \hfil \textbf{Research Question} & \hfil \textbf{Motivation} \\ \hline
         \textbf{RQ1:} What are the visualization and interaction techniques used by ICSA systems? & To identify visualization techniques and interaction features used by immersive technologies for cyber SA. \\ \hline
         \textbf{RQ2:} What level of cyber SA is facilitated by immersive technologies in ICSA systems? & To identify perception, comprehension, and projection of cyber SA in ICSA systems. \\ \hline
         \textbf{RQ3:} How are ICSA systems evaluated? & To identify evaluation techniques for validating ICSA systems. \\ \hline
    \end{tabular}
    \par}
\end{center}
\end{table*}

\subsection{Search Strategy}

We devised a comprehensive search strategy for extracting as many relevant studies as possible from the Scopus search engine. The search strategy was composed of the following steps. 

\begin{itemize}[leftmargin=*]

\item \textit{Search Method:} We first designed an inclusive search string containing the terms related to our research questions. Then, we ran the search string on Scopus to retrieve the maximum relevant studies on ICSA systems. 

\item \textit{Search Terms:} Our search string included all the terms that are relevant to the research objectives (i.e., RQs) of this survey. \textcolor{blue} {Fig.~\ref{fig:string}} shows the developed search string that is mainly composed of two parts: (i) the first part consisted of different \textit{"immersive technologies"}, and (ii) the second part contained the synonyms and relevant terms of \textit{"cybersecurity"} and \textit{"cyber situational awareness"}. It is important to note that the terms were searched in the title, keywords, and abstract of the papers available at Scopus to identify and extract the relevant literature on ICSA systems.

\begin{figure} 
  \centering
  \includegraphics[width=0.98\columnwidth]{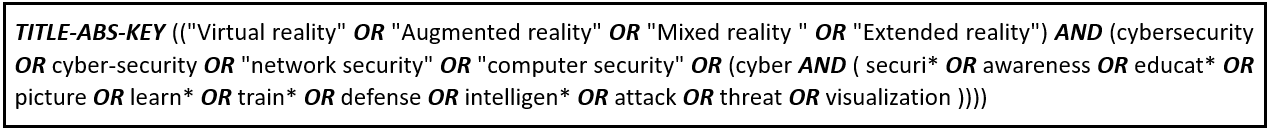}
  \caption{Search string for this survey.}
  \label{fig:string} 
  \vspace{-15pt}
\end{figure}

\item \textit{Data Sources:} Similar to \cite{dissanayake2022software}, we used the Scopus search engine only to identify the relevant literature on ICSA systems for this survey. This is mainly because of the observations reported in \cite{shahin2017continuous, kitchenham2010systematic, zahedi2016systematic, shahin2020architectural} that justify Scopus indexes a large amount of peer-reviewed papers and journals indexed by many other digital databases such as IEEE Xplore, ACM Digital Library, Science Direct, SpringerLink and Wiley Online Library.

\end{itemize}

\subsection{Inclusion and Exclusion Criteria}

We devised inclusion and exclusion criteria to select the most relevant studies on ICSA systems for this survey. We included the peer-reviewed papers that can answer our defined research questions. Moreover, we included the study that is written in English language only, irrespective of its publication date. All other types of literature, such as workshop articles, editorials, keynotes, tutorial summaries, panel discussions, books, and reviews, are out of the scope of this survey. \textcolor{blue} {Table~\ref{tab:criteria}} presents the inclusion and exclusion criteria of this survey. The criterion was applied in the study selection process to retrieve the most pertinent papers, as described in the next phase.

\renewcommand{\arraystretch}{1.5}
\begin{table}[t]
\captionof{table}{Inclusion and exclusion criteria for this survey.} 
\label{tab:criteria} 
\begin{center}
    {\fontsize{9}{11}\selectfont
     \begin{tabular}{|m {45em}|}
         \hline
         \textbf{Inclusion Criteria} \\ \hline
         \textbf{\textit{I1:}}~ A study that is related to the use of immersive technologies for cybersecurity purposes \\ \hline
          \textbf{\textit{I2:}}~ A study is selected irrespective of its publication date \\ \hline
           \textbf{Exclusion Criteria} \\ \hline
         \textbf{\textit{E1:}}~ A study that is written in a language other than English \\ \hline
         \textbf{\textit{E2:}}~ Books, workshop articles and non-peer-reviewed papers \\ \hline
         \textbf{\textit{E3:}}~ Full text is not accessible \\ \hline
   \end{tabular}
   \par}
\end{center}
\end{table}

\vspace{-10pt}

\subsection{Study Selection}

We identified, selected, and extracted the existing state-of-the-art ICSA systems through a six-step process. The phases of the study selection process are briefly described as follows:

\begin{itemize}[leftmargin=*]

    \item \textit{Automatic Search:} We ran our search string on the Scopus search engine to identify existing literature on ICSA systems. As a result, we retrieved 3536 potential studies.
    
    \item \textit{Title-based Selection:} We analyzed the title of the 3536 studies. If a paper title was relevant to the research questions of this survey, we included that paper. In case when we were not sure about the relevance of a paper, that paper was transferred to the next phase. At the end of this phase, we had 327 papers.

    \item \textit{Duplication Removal:} As we consulted one database (i.e., Scopus) only to retrieve the existing literature, no duplicate study was found during our study selection process.

    \item \textit{Abstract-based selection:} We thoroughly read the abstracts and conclusions of the remaining 327 studies to check their relevance to our research questions. Here, we also applied the inclusion and exclusion criteria (\textcolor{blue} {Table~\ref{tab:criteria}}) to the abstracts of papers. Consequently, this phase reduced the pool of papers from 327 to 139.

    \item \textit{Full-text based selection:} We read the full text of 139 studies, and applied the inclusion and exclusion criteria on them. As a result, we got 36 relevant studies.

    \item \textit{Snowballing:} We performed forward and backward snowballing \cite{wohlin2014guidelines} on the 36 studies to identify more literature on ICSA systems. This gave us 7 potentially relevant studies that were then passed through our inclusion and exclusion criteria. Consequently, we finalized 43 relevant studies for this survey.

\end{itemize}

\vspace{-10pt}

\subsection{Data Extraction and Synthesis}

In this section, we report the data extraction and synthesis process of this survey. First, we extracted the data relevant to our research questions from the finalized 43 studies. Then, we analyzed and synthesized the extracted data to answer the research questions of this survey. The details of each process are as follows.

\subsubsection{Data Extraction}

We formulated a list of data items according to our research questions for extracting relevant details from the retrieved studies. \textcolor{blue} {Table~\ref{tab:extraction}} presents the data extraction form containing the list of data items prepared for this survey. The data items D1 to D8 present demographic details of the extracted literature. For example, title (D1), author(s) (D2), and publication type (D5) of papers represent the demographic data of the existing literature. Similarly, the data items D9 to D12 correspond to our research questions. For example, the details of visualization techniques and interaction features were collected against data items D9 and D10, respectively, to answer RQ1. We papered a Microsoft Excel spreadsheet to save the extracted information against each data item for further analysis.

\renewcommand{\arraystretch}{1.2}
\begin{table}[t]
\captionof{table}{Data extraction form of this survey.} 
\label{tab:extraction} 
\begin{center}
    {\fontsize{9}{11}\selectfont
     \begin{tabular}{|C{2em}|C{6em}|C{25em}|C{10em}|}
         \hline
         \textbf{ID} &  \textbf{Data Item} &  \textbf{Description} &  \textbf{Research Questions} \\ \hline
             D1	&  Title	&  The title of the paper	& Demographic data \\ \hline
             D2	&  Author(s)	&  The author(s) of the paper	& Demographic data \\ \hline
             D3	&  Venue	&  The publication venue	& Demographic data \\ \hline
             D4	&  Year	&  The year of the publication	& Demographic data \\ \hline
             D5	&  Publication type & The type of publication (e.g., journal paper, conference paper) & Demographic data \\ \hline
             D6	&  Area of focus	&  The focus of the paper in ICSA domain & Demographic data \\ \hline 
             D7	&  Target user(s) &  The intended user(s) (e.g., security analyst) & Demographic data \\ \hline
             D8	&  Software and hardware tools &  The software/hardware tools used for ICSA systems & Demographic data \\ \hline
             D9	& Visualization techniques	& The visualization techniques for ICSA systems & RQ1 \\ \hline 
             D10	&  Interaction techniques & The interaction techniques for ICSA systems & RQ1 \\ \hline 
             D11 & Cyber SA level	& The level of cyber SA achieved for ICSA systems &  RQ2 \\ \hline 
             D12	&  Evaluation mechanisms & The evaluation mechanisms used to validate ICSA systems & RQ3 \\ \hline
             D13	& Future work  &  The reported future work & Discussion \\ \hline
             
   \end{tabular}
   \par}
\end{center}
\end{table}

\subsubsection{Data Synthesis}

We used descriptive statistics to analyze the demographic attributes (i.e., D1 to D8) while the other data items (i.e., D9 to D13) were analyzed by using thematic analysis \cite{braun2006using}. The thematic analysis methodology identifies themes in the extracted data, interprets the themes, and draws conclusions. As per the thematic analysis guidelines reported by Braun and Clarke \cite{braun2006using}, we followed the five-step approach to perform thematic analysis on the data items D9 to D13.

\begin{itemize}[leftmargin=*] 

\item \textit{Familiarizing with data:} We got an initial understanding of the extracted visualization techniques (D9), interaction features (D10), levels of cyber SA (D11) and evaluation mechanisms (D12) for ICSA systems. 

\item \textit{Generating initial codes:} We developed a rudimentary list of similar visualization techniques, interaction features, cyber SA levels, and evaluation mechanisms for ICSA systems. In some cases, we re-examined the retrieved studies to verify the developed list.

\item \textit{Searching for themes:} We categorized the initial codes for each data item into potential themes. For example, visualization techniques based on icons are combined under the theme of "Iconic Displays".

\item \textit{Reviewing and refining themes:} We analyzed the identified themes against each other to detect similar and irrelevant themes. For example, spatial visualizations and geometric visualizations were merged with each other due to their same characteristics.

\item \textit{Defining and naming themes:} A clear and concise name was defined for each theme.

\end{itemize}



\section{Visualization and Interaction Techniques}  \label{visualization&Interactions}
This section addresses \textit{RQ1: What are the visualization and interaction techniques used by ICSA systems?} It examines the specific visualization techniques and interaction features employed by immersive technologies to enhance cyber SA. Section \ref{visualization} discusses the visualization techniques, while Section \ref{Interactions} outlines the interaction features used in ICSA systems.

\renewcommand{\arraystretch}{1.2}
\begin{table}[b]
\captionof{table}{Identified immersive visualization techniques and their sources.} 
\label{tab:Visualization} 
\begin{center}
    {\fontsize{9}{11}\selectfont
     \begin{tabular}{|C{15em}|C{30em}|}
         \hline
         \textbf{Visualization Techniques} &  \textbf{Papers} \\ \hline
         
             Geographical Displays	& \cite{mattina2017marcs}, \cite{sukhija2019employing}, \cite{beitzel2017exploring}, \cite{ma2018learning}  \\ \hline
             
             Metaphorical Displays	& \cite{delcombel2021cybercopter} \\ \hline
             
             Node-Link Graphs	& \cite{beitzel2017exploring}, \cite{kabil20183d}, \cite{kabil2020alert}, \cite{kullman2019operator}, \cite{ask20233d}  \\ \hline
             
             Scatterplots & \cite{kullman2018enhancing} \\ \hline
             
             3D Bar Charts & \cite{beitzel2018network} \\ \hline 
             
             Volume	& \cite{sukhija2019employing}, \cite{korkiakoski2021using}, \cite{beitzel2017exploring}, \cite{alqahtani2020design}, \cite{beitzel2016cognitive}, \cite{seo2019using}, \cite{chu2019data}, \cite{jin2018game}, \cite{kabil20183d}, \cite{kabil2020alert}, \cite{salazar2013enhancing}, \cite{delcombel2021cybercopter}, \cite{garae2017visualizing}, \cite{puttawong2017vrfiwall}, \cite{kasurinen2017usability},\cite{sharma2019security}  \\ \hline
             
             Icons/Symbols/Glyphs	& \cite{mattina2017marcs}, \cite{beitzel2017exploring}, \cite{chiou2021augmented}, \cite{jin2018game}, \cite{salazar2013enhancing}, \cite{sharma2019security}, \cite{Kaleem2019security}  \\ \hline
             
             Animation/Video Displays & \cite{sukhija2019employing}, \cite{chiou2021augmented}, \cite{Kaleem2019security} \\ \hline
             
             360$^{\circ}$ Pictures & \cite{rana2014exploring}, \cite{sharma2019security}   \\ \hline
             
             Two-Dimensional Displays	& \cite{korkiakoski2021using}, \cite{beitzel2017exploring}, \cite{alqahtani2020design}, \cite{beitzel2016cognitive}, \cite{seo2019using}, \cite{chu2019data}, \cite{kabil20183d}, \cite{kabil2020alert}, \cite{Kaleem2019security}  \\ \hline
             
             List/Table/Text Displays & \cite{mattina2017marcs}, \cite{sukhija2019employing}, \cite{alqahtani2020design}, \cite{seo2019using}, \cite{chu2019data}, \cite{chiou2021augmented}, \cite{kabil2020alert}, \cite{delcombel2021cybercopter}, \cite{sharma2019security}, \cite{Kaleem2019security}  \\ \hline

   \end{tabular}
   \par}
\end{center}
\end{table}

\subsection{Visualization Techniques for ICSA} \label{visualization}

Immersive technologies provide several unique features for complex and multi-dimensional cybersecurity data visualization to enhance the understanding of end-users. For example, these features include spatial immersion, situated analytics, embodied data exploration, collaboration, multi-sensory presentation, and engagement for complex data presentation and analysis \cite{dwyer2018immersive}. This enhances the perception, comprehension, and projection of cyber SA for end-users. To answer RQ1, we have identified 11 distinguish visualization techniques of immersive technologies, from the existing literature, used for creating cyber SA. \textcolor{blue} {Table~\ref{tab:Visualization}} presents the identified visualization techniques along with their corresponding literature. In the following, we report the details of each visualization technique for ICSA.

\begin{figure}[t]
\minipage{0.32\textwidth}
  \includegraphics[width=\linewidth, height=0.2\textheight]{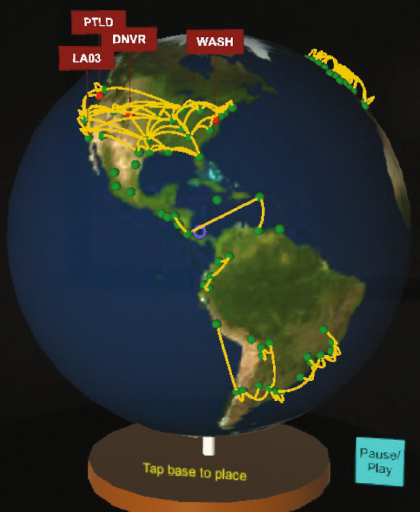}
  \caption{Geographical display \cite{beitzel2017exploring}.}\label{fig:geographical}
\endminipage\hfill\minipage{0.32\textwidth}
  \includegraphics[width=\linewidth, height=0.2\textheight]{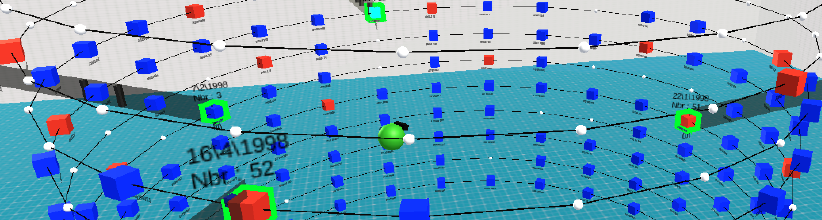}
  \caption{Metaphorical display \cite{delcombel2021cybercopter}.}\label{fig:metaphorical}
\endminipage\hfill\minipage{0.32\textwidth}%
  \includegraphics[width=\linewidth, height=0.2\textheight]{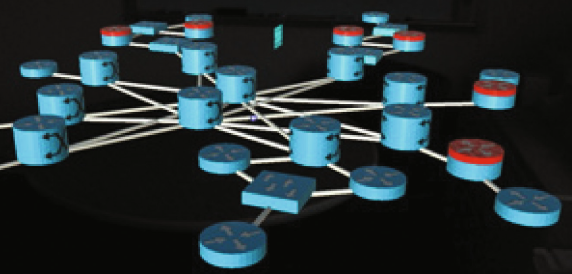}
  \caption{Node-link graph \cite{beitzel2017exploring}.}\label{fig:nodelink}
\endminipage\end{figure}

\textbf{Geographical Displays:} ICSA systems use geographical displays to represent cybersecurity data for creating cyber SA. Geographical displays refer to the visualizations that present cybersecurity data with their geographical locations (e.g., position coordinates of longitude and latitude). The spatial information helps end-users (e.g., cybersecurity experts, trainees, and analysts) in understanding a cyber situation, which facilitates their decision-making and execution of action plans accordingly. For instance, an immersive 3D network visualizer application, developed by Beitzel et al. \cite{beitzel2017exploring}, displays networks' topology with their geographic coordinates. In each network topology, network nodes, and edges are overlaid onto a real-world globe based on their geographical locations, as presented by \textcolor{blue} {Fig.~\ref{fig:geographical}}. It can also be seen that cybersecurity alerts/flags are placed on corresponding nodes and edges, which facilitates end-users in monitoring and detecting cybersecurity anomalies in large-scale networks.

\textbf{Metaphorical Displays:} Cybersecurity visualizations leverage metaphors to represent cyber situations in a comprehensible manner. Metaphorical displays provide clarity and context of cyber situations without any data occlusion. For example, Delcombel et al. \cite{delcombel2021cybercopter} developed a helix structure to display arranged, organized, and systematized cybersecurity data. The 3D helical representation, presented by \textcolor{blue} {Fig.~\ref{fig:metaphorical}}, helps users in monitoring and detecting periodic signals of cyber-attacks. A user is placed inside the helix, surrounded by cybersecurity data, to get insights into cyber situations with proper context and less obstruction.

\textbf{Node-Link Graphs:} Large-scaled network topologies are presented by node-link graphs to identify and understand how nodes (e.g., entities) are connected with each other. This provides an overview of network topologies, where nodes and their links are drawn as points and lines respectively. For instance, the 3D Cyber COP prototype, developed by Kabil et al. \cite{kabil20183d}, presents a holistic view of network topologies through node-link graphs for cybersecurity coordinators, which provide them an understanding of network security state. Similarly, the 3D Network Visualizer application \cite{beitzel2017exploring} contracts node-link graphs for showing network assets and their relationships with each other. \textcolor{blue} {Fig.~\ref{fig:nodelink}} shows the node-link graphs developed by the 3D Network Visualizer application for cyber situational awareness.

\begin{figure}[t]
\minipage{0.32\textwidth}
  \includegraphics[width=\linewidth, height=0.2\textheight]{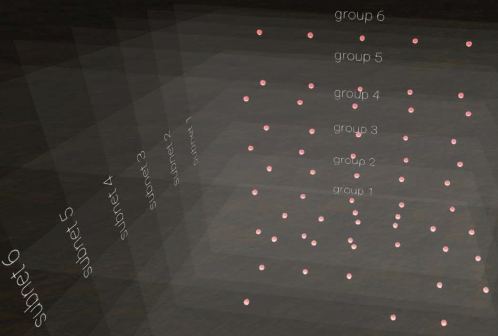}
  \caption{Scatterplot \cite{kullman2018enhancing}.}\label{fig:scatterplots}
\endminipage\hfill\minipage{0.32\textwidth}
  \includegraphics[width=\linewidth, height=0.2\textheight]{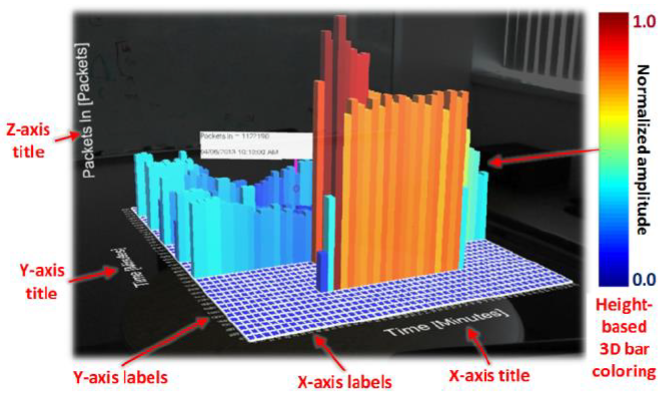}
  \caption{3D bar chart \cite{beitzel2018network}.}\label{fig:barchart}
\endminipage\hfill\minipage{0.32\textwidth}%
  \includegraphics[width=\linewidth, height=0.2\textheight]{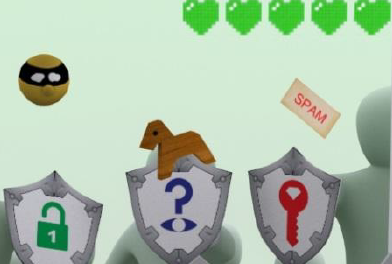}
  \caption{Volume \cite{salazar2013enhancing}.}\label{fig:volume}
\endminipage\end{figure}

\textbf{Scatterplots:} Cybersecurity data need to be correlated with different cybersecurity parameters for a better understanding of cyber situations. For this purpose, immersive three-dimensional scatterplots are used to display relationships among cybersecurity data. Moreover, additional cybersecurity information can be visualized through different colors, shapes, and sizes of objects in scatterplots. An interesting example of immersive scatterplots is reported in \cite{kullman2018enhancing} where different sets of network traffic data are displayed through scatterplots for different networks (\textcolor{blue} {Fig.~\ref{fig:scatterplots}}). The color, shape, and size of data objects present different cybersecurity parameters (e.g., anomalous and normal traffic data) for better perception and analysis of cyber situations.

\textbf{3D Bar Charts:} 3D bar charts refer to the representation of information with rectangular bars of different widths, heights, and colors presenting the corresponding cybersecurity data in an immersive three-dimensional space. These charts are mostly used to perform a comparative analysis of multi-dimensional data in the cybersecurity realm. For instance, Beitzel et al. \cite{beitzel2018network} developed a prototype application, called MINER, to perform a network anomaly analysis with 3D bar charts (\textcolor{blue} {Fig.~\ref{fig:barchart}}). The unique design of these 3D bar charts allows for a more intuitive and time-efficient identification of irregularities within network data. By presenting the information in this visually accessible manner, MINER enhances the cyber SA of its end-users.

\textbf{Volume:} Volume refers to the 3D object representations of cybersecurity data, encompassing assets, cyber-attacks, and countermeasures, to aid end-users in understanding and managing cyber situations. For example, \textcolor{blue} {Fig.~\ref{fig:volume}} presents the cartoonish 3D models and virtual shields that are used to show cybersecurity threats and countermeasures respectively in an AR-based serious game reported in \cite{salazar2013enhancing}. Similarly, 3D round-shaped objects (e.g., spheres), of red and green colors to indicate faulty and normal equipment respectively, are used in debug UIs developed to detect anomalies in high-performance computing environments \cite{sukhija2019employing}. The literature reporting the use of \textit{volume} in ICSA systems is presented in \textcolor{blue} {Table~\ref{tab:Visualization}}.

\begin{figure}[t]
\minipage{0.32\textwidth}
  \includegraphics[width=\linewidth, height=0.2\textheight]{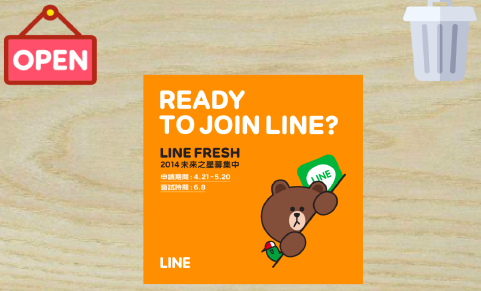}
  \caption{Icons \cite{chiou2021augmented}.}\label{fig:icon}
\endminipage\hfill\minipage{0.32\textwidth}
  \includegraphics[width=\linewidth, height=0.2\textheight]{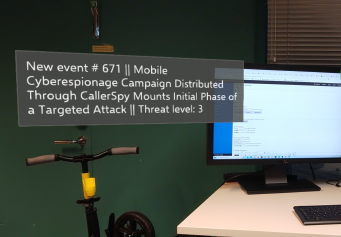}
  \caption{2D display \cite{korkiakoski2021using}.}\label{fig:2display}
\endminipage\hfill\minipage{0.32\textwidth}%
  \includegraphics[width=\linewidth, height=0.2\textheight]{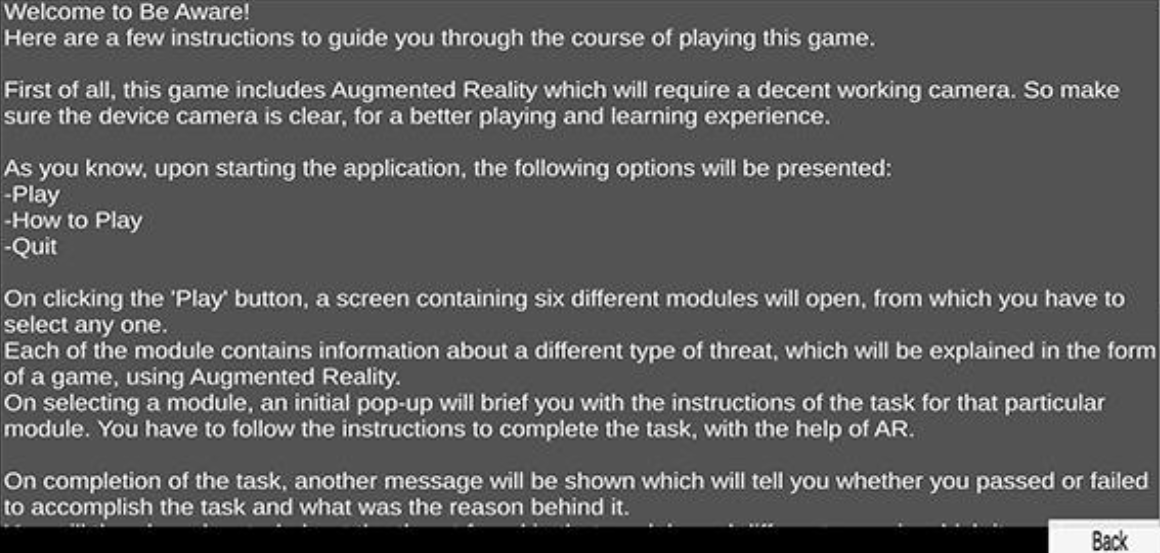}
  \caption{Text Display \cite{sharma2019security}.}\label{fig:textual}
\endminipage\end{figure}

\textbf{Icons/Symbols/Glyphs:} ICSA systems use icons, symbols, or glyphs to represent cybersecurity data for better perception and comprehension of end-users. Our survey shows that this immersive visualization technique is extensively used in ICSA systems for cyber SA (\textcolor{blue} {Table~\ref{tab:Visualization}}). For instance, an AR-based application, reported in \cite{chiou2021augmented}, uses icons to indicate objects for navigating into the application, such as \textit{trash-bin icon} for delete and \textit{open icon} for open operation, as presented by \textcolor{blue} {Fig.~\ref{fig:icon}}. Moreover, different other icons, such as email and trashcan icons, are used to open and delete an email. Similarly, the 3D Network Visualizer applications use icons to show a variety of network components, including switches, computers, and routers \cite{beitzel2017exploring}. The use of icons, symbols, or glyphs increases the understanding and perception of cyber situations of end-users.

\textbf{Animation/Video Displays:} ICSA systems use animations and videos to display cybersecurity information (e.g., impacts of cyber-attacks, and countermeasures' implementation process) for enhancing cyber SA of end-users. For example, Sukhija et al. \cite{sukhija2019employing} presented a concept of animation in their developed AR-based framework for fixing anomalies in large-scale infrastructures. For instance, animations include instructions to fix an anomaly to enhance cybersecurity. Similarly, the application on phishing education, developed by Chiou et al.\cite{chiou2021augmented}, uses animations to demonstrate the impacts (e.g., the disappearance of applications, photos, or emails) of phishing attacks. This sort of information enables end-users to predict cyber situations caused by their actions.

\textbf{360$^{\circ}$ Pictures:} Immersive visualizations employ 360$^{\circ}$ pictures to disseminate cyber SA to end-users. 360$^{\circ}$ pictures present an overall and comprehensive view of cyber situations, which enhances the cyber SA of end-users. A potential use of 360$^{\circ}$ pictures is highlighted in \cite{rana2014exploring} in which 360$^{\circ}$ pictures are used to detect cybersecurity vulnerabilities in a given system. The identification of cybersecurity vulnerabilities indicates the severity of a cyber situation. It also leads to the identification of potential countermeasures for eradicating the detected vulnerabilities.

\textbf{Two-Dimensional Displays:} ICSA systems use two-dimensional displays for cybersecurity 2D data visualization in a 3D spatial space around end-users. For example, focusing on a single display of information, Korkiakoski et al. \cite{korkiakoski2021using} developed an AR application that shows cybersecurity data (e.g., threats and severity levels) through a virtual 2D event information panel along with physical monitors that show operational information, as presented by \textcolor{blue} {Fig.~\ref{fig:2display}}. Similarly, the node-link graphs, presented in \cite{beitzel2017exploring}, display network data (i.e., command terminal windows and Wireshark displays) through 2D displays in an immersive 3D space. Two-dimensional displays are usually integrated with 3D visualization techniques in ICSA systems to enhance the cyber SA of end-users.

\textbf{List/Table/Text Displays:} ICSA systems need to display cybersecurity information through long-text, list or tabular format. Therefore, immersive technologies employ lists, tables, or text summaries to represent cybersecurity information for enhancing the cyber SA of end-users. For example, a VR system, developed in \cite{seo2019using}, provides textual instructions for entering, navigating, and inspecting a virtual data center through a virtual tablet. Moreover, operational and security protocols are provided to end-users through the tablet for detecting and fixing an anomaly in the data center. Similarly, an AR-based cybersecurity awareness game \cite{sharma2019security} provides textual instructions to end-users regarding how to play the game (\textcolor{blue} {Fig.~\ref{fig:textual}}); also, it gives an explanation behind each step taken by users in a textual format to raise their cybersecurity awareness.

\renewcommand{\arraystretch}{1.2}

\begin{table}[t]
\captionof{table}{Identified immersive interaction techniques and their sources.}
\label{tab:Interaction} 
\begin{center}
    {\fontsize{9}{11}\selectfont
     \begin{tabular}{|C{15em}|C{30em}|}
         \hline
         \textbf{Interaction Techniques} &  \textbf{Papers} \\ \hline
         
             Select	& \cite{mattina2017marcs}, \cite{sukhija2019employing}, \cite{beitzel2017exploring}, \cite{alqahtani2020design}, \cite{seo2019using}, \cite{chu2019data}, \cite{chiou2021augmented}, \cite{kabil2018should}, \cite{kabil20183d}, \cite{kabil2020alert}, \cite{salazar2013enhancing}, \cite{delcombel2021cybercopter}, \cite{ma2018learning}, \cite{puttawong2017vrfiwall}, \cite{sharma2019security}, \cite{Kaleem2019security}, \cite{beitzel2018network}, \cite{kullman2018enhancing}, \cite{korkiakoski2023hack} \\ \hline
             
             Navigate	& \cite{sukhija2019employing}, \cite{beitzel2017exploring}, \cite{alqahtani2020design}, \cite{seo2019using}, \cite{chu2019data}, \cite{chiou2021augmented}, \cite{kabil2018should}, \cite{kabil20183d}, \cite{kabil2020alert}, \cite{delcombel2021cybercopter}, \cite{ma2018learning}, \cite{puttawong2017vrfiwall}, \cite{Kaleem2019security}  \\ \hline
             
             Details on Demand	& \cite{mattina2017marcs}, \cite{sukhija2019employing}, \cite{beitzel2017exploring}, \cite{alqahtani2020design}, \cite{seo2019using}, \cite{chu2019data}, \cite{chiou2021augmented}, \cite{kabil20183d}, \cite{kabil2020alert},
             \cite{chakal2023augmented}, \cite{delcombel2021cybercopter}, \cite{ma2018learning}, \cite{puttawong2017vrfiwall}, \cite{sharma2019security}, \cite{Kaleem2019security}, \cite{beitzel2018network}, \cite{bernsland2022cs}  \\ \hline
             
             Arrange/Change & \cite{beitzel2017exploring}, \cite{seo2019using}, \cite{chu2019data}, \cite{chiou2021augmented}, \cite{kabil20183d}, \cite{kabil2020alert}, \cite{delcombel2021cybercopter}, \cite{ma2018learning}, \cite{beitzel2018network}, \cite{kullman2018enhancing} \\ \hline
             
             Filter	& \cite{beitzel2017exploring}, \cite{kabil20183d}, \cite{delcombel2021cybercopter} \\ \hline
             
             Extract/Share	& \cite{kabil2018should}, \cite{kabil20183d}  \\ \hline
             
             Aggregate/Relate &  \cite{kabil20183d}, \cite{delcombel2021cybercopter}, \cite{beitzel2018network}, \cite{kullman2018enhancing} \\ \hline
             
             Annotate & \cite{kabil20183d}, \cite{delcombel2021cybercopter} \\ \hline
             
             Record & \cite{kabil2020alert} \\ \hline 
        
   \end{tabular}
   \par}
\end{center}
\end{table}

\subsection{Interaction Techniques for ICSA} \label{Interactions}

An interaction technique refers to an approach to interact with immersive visualizations for creating and maintaining cyber SA for a given system. We have identified 9 interaction techniques for ICSA systems through the existing literature. \textcolor{blue} {Table~\ref{tab:Interaction}} presents the identified interaction techniques with their corresponding literature. In the following, we describe each interaction technique of ICSA systems for creating cyber SA.

\textbf{Select:} Users interact with ICSA systems by selecting objects/options to perform different activities (e.g., arranging and manipulating cybersecurity data) for getting cyber SA. Immersive technologies provide several ways for the selection task in ICSA systems. This includes touch-selection \cite{mattina2017marcs}, gaze-selection \cite{beitzel2018network}, point-selection \cite{puttawong2017vrfiwall}, gesture-selection \cite{beitzel2017exploring}, controller-selection \cite{seo2019using}, and custom marker selection \cite{salazar2013enhancing}. These natural interactions make the selection task easier for end-users than the traditional mouse and keyboard selection in which investigation of large-scale networks becomes cumbersome when users try to reach a specific node \cite{kabil2018should}.

\textbf{Navigate:} Navigation refers to the interactions that help users move in ICSA systems to get cybersecurity awareness. Most immersive technologies employ head tracking to enable their users to navigate ICSA systems with physical movements \cite{puttawong2017vrfiwall}. Head tracking provides a natural experience of moving around 3D virtual elements for getting a comprehensive perception and comprehension of cyber SA \cite{kabil2018should}. Similarly, immersive technologies use selection techniques with virtual objects, allowing users to navigate ICSA systems. For example, users point and click on 3D arrows to navigate around a VR environment presented in \cite{seo2019using}. Another navigation technique is zooming which provides users zoom in and out capabilities for navigating ICSA environments.

\textbf{Details on Demand:} Users of ICSA systems need necessary details of cybersecurity data for understanding, analyzing, and forecasting cyber situations. Therefore, immersive technologies offer \textit{details on demand} capability in ICSA systems to display detailed information about cybersecurity data, when required by end-users. For instance, the 3D Network Visualizer application shows the detailed network traffic, through command terminal windows and Wireshark displays, on top of nodes in a node-link graph to diagnose a network cyber situation \cite{beitzel2017exploring}.

\textbf{Arrange/Change:} Organization of cybersecurity data provides insights of cyber situations to end-users. Therefore, ICSA systems enable users to arrange and change cybersecurity data, statistics, and views in a comprehensible manner so that they can perceive cyber situations with minimum cognitive load. This includes information highlighting, changing attribute mapping, and changing representations (i.e., customization) of cybersecurity visualizations, which enhances users' cyber SA. For example, users can change angles between radial data and helix of the helical visualization, proposed in \cite{delcombel2021cybercopter}, to visualize cybersecurity data with clarity.

\textbf{Filter:} Filters enable users to apply inclusion and exclusion criteria on cybersecurity visualization elements to maintain their focus on essential cybersecurity data for getting cyber SA. For instance, the Network Feed application, proposed in \cite{beitzel2017exploring}, allows users to filter network traffic types (e.g., UDP, ICMP, and TCP) to get specific insights into cyber situations.

\textbf{Extract/Share:} ICSA systems allow users to extract and share cybersecurity reports, status, and visualizations with each other for creating collaborative environments to estimate cyber SA. Moreover, the collaborative assessment of cyber SA facilitates the identification, selection, and preparation of an optimal action plan for achieving desired cyber situations. For instance, the 3D Cyber COP prototype, developed by Kabil et al. \cite{kabil20183d}, shares different cybersecurity data visualizations with operators according to their roles (e.g., analyst, coordinator, and client). It also provides cybersecurity report extraction capability to coordinators at any given instance to perform collaborative analysis for estimating cyber situations in real time.

\textbf{Aggregate/Relate:} ICSA systems enable users to aggregate cybersecurity data to make sense of what is going on in cyberspace. Recalling the example of the 3D Cyber COP prototype, cybersecurity experts combine raw data to create potential cyber incident scenarios \cite{kabil20183d}. This helps users predict possible cyber situations and prepare possible action plans accordingly.

\textbf{Annotate:} Annotation refers to the addition of graphical or textual information on cybersecurity visualizations for a better understanding of cyber SA. The metaphorical display of cybersecurity information allows users to add additional information on a given space beyond the helix \cite{delcombel2021cybercopter}. Similarly, the 3D Cyber COP application enables cybersecurity analysts to share their analysis with each other by adding visual cues on cyber assets \cite{kabil20183d}. In this way, all the stakeholders involved in cybersecurity operations can get the same picture of cyberspace, which facilitates the execution of cybersecurity operations in a collaborative manner.

\textbf{Record:} ICSA systems allow users to save their interaction logs and cybersecurity data trends for estimating the cyber situations of a given system. Historical data help users in detecting anomalies and predicting cyber situations. For example, the 3D Cyber COP application shows a system's parameter details (e.g., status, trend, and history) through a two-dimensional graph on a virtual 2D screen to help analysts in assessing cyber SA \cite{kabil2020alert}.



\section{Levels of Immersive Cyber Situational Awareness}  \label{ICSA_Levels}

This section answers \textit{RQ2, what level of cyber SA is facilitated by immersive technologies in ICSA systems?} Given the wide acceptance of Endsley's SA model \cite{endsley1988design} in several domains (e.g., paramedicine \cite{hunter2020towards}, military realm \cite{riley2006collaborative} and cybersecurity \cite{onwubiko2016understanding}), we have leveraged the Endsley's SA model for assessing cyber SA in immersive environments. Accordingly, we have defined and identified the three levels (i.e., perception, comprehension, and projection) of SA for ICSA through the existing literature. \textcolor{blue} {Table~\ref{tab:ICSA_Levels}} presents the ICSA levels with their corresponding literature. In the following, we describe the definition and details for each ICSA level.

\subsection{Perception} \label{perception}

Perception is the fundamental phase of SA, which enables users to answer the question \textit{what is happening in cyber environments?} For immersive environments, perception refers to the monitoring, detecting, and recognizing cybersecurity data (e.g., attack vectors, vulnerabilities, and countermeasures) that provide users with a holistic picture of cyber situations. Immersive visualization and interaction techniques help users create and maintain their perception of cyber SA in ICSA systems. For instance, the immersive display, proposed in \cite{korkiakoski2021using}, shows an overview of ongoing cyber threats with their severity levels for COVID-19 information systems, which provides security experts an overall perception of cyber situations. Similarly, the VR-based cybersecurity game, developed by Jin et al. \cite{jin2018game}, uses icons for cyber-attacks and defenses to help users recognize cybersecurity elements, which enhances the perception of end-users for cyber SA. \textcolor{blue} {Table~\ref{tab:ICSA_Levels}} presents the reviewed studies that address the perception level of cyber SA for ICSA systems.

\renewcommand{\arraystretch}{1.2}
\begin{table}[t]
\captionof{table}{Identified levels of ICSA from existing literature.} 
\label{tab:ICSA_Levels} 
\begin{center}
    {\fontsize{9}{11}\selectfont
     \begin{tabular}{|C{15em}|C{30em}|}
         \hline
         \textbf{Levels of ICSA} &  \textbf{Papers} \\ \hline
         
             Perception	& \cite{mattina2017marcs}, \cite{sukhija2019employing}, \cite{korkiakoski2021using}, \cite{alqahtani2020exploring}, \cite{shen2021work}, \cite{beitzel2016cognitive}, \cite{seo2019using}, \cite{chu2019data}, \cite{wagner2023leveraging}, \cite{chiou2021augmented}, \cite{jin2018game}, \cite{kabil20183d}, \cite{kabil2020alert}, \cite{salazar2013enhancing}, \cite{delcombel2021cybercopter}, \cite{kommera2016smart}, \cite{garae2017visualizing}, \cite{ma2018learning}, \cite{puttawong2017vrfiwall}, \cite{kasurinen2017usability}, \cite{sharma2019security}, \cite{Kaleem2019security}, \cite{beitzel2018network}, \cite{kullman2018enhancing}, \cite{korkiakoski2023hack}, \cite{chakal2023augmented}, \cite{chen2021exploring} \\ \hline
             
             Comprehension	& \cite{beitzel2017exploring}, \cite{alqahtani2020design}, \cite{kabil20183d} \cite{chiou2021augmented}, \cite{kullman2019vr}, \cite{kabil2020alert}, \cite{delcombel2021cybercopter}, \cite{beitzel2018network}, \cite{veneruso2020cybervr}, \cite{kullman2019operator}, \cite{bernsland2022cs}, \cite{kaneko2020packuarium}  \\ \hline
             
             Projection	& \cite{sukhija2019employing}, \cite{alqahtani2020design}, \cite{chiou2021augmented}, \cite{ask20233d}   \\ \hline
        
   \end{tabular}
   \par}
\end{center}
\end{table}

\subsection{Comprehension} \label{comprehension}

Though the perception level of SA provides a basic understanding of cyber situations, comprehension conveys extensive knowledge about cyberspace to end-users. The comprehension level of SA enables users to answer the questions \textit{"Why is it happening?" and “What is its meaning?”} ICSA systems enable users to explore, analyze, and investigate cyber situations through interactive visualizations. For example, the 3D Cyber COP model allows non-experts to distinguish between malicious alerts and false positives through different visualization and interaction techniques \cite{kabil2020alert}. Similarly, Beitzel et al. \cite{beitzel2018network} presents interactive bar charts for comparative analysis to detect anomalies in network traffic. The comprehension level of ICSA covers data analysis tasks that include, but are not limited to, data clustering, anomaly detection, pattern analysis, visual search, comparative analysis, and data enrichment. \textcolor{blue} {Table~\ref{tab:ICSA_Levels}} presents the literature that reports the use of immersive technologies to enhance cyber SA.



\subsection{Projection} \label{projection}

Projection refers to the prediction of cyber situations, which helps users in answering the questions \textit{“What will happen next?” and “What can I do?”} Immersive technologies allow users to envisage the evolution of cyber situations with less cognitive load and mental stress. For example, the AR-based cybersecurity education application shows the impacts of phishing attacks (e.g., the disappearance of photos, apps, or emails) \textit{animations} when users make wrong choices \cite{chiou2021augmented}. This helps users predict the potential consequences of phishing attacks when users operate in real-world scenarios. We have identified a few studies, as presented by \textcolor{blue} {Table~\ref{tab:ICSA_Levels}}, that describe the projection of cyber situations using immersive technologies. These studies underscore the importance of the projection phase in ICSA systems, highlighting how visualization and interaction techniques can aid users in anticipating and mitigating cyber threats.

\section{Evaluation Mechanisms for ICSA Systems} \label{evaluation}

This section answers \textit{RQ3: How are ICSA systems evaluated?} From the reviewed studies, we identified that researchers have employed various user-experience research methods to assess ICSA systems. These methods include questionnaires, surveys, situational awareness evaluation techniques, and usability evaluation mechanisms. 

User-oriented studies have provided comprehensive insights into several key aspects. They report on users' demographics, including factors such as gender and ethnicity, which help in understanding the diversity of participants involved in the studies. Performance metrics such as threat response time and task completion time are also analyzed, offering quantitative measures of user efficiency and effectiveness when interacting with ICSA systems. Additionally, cognition parameters such as distraction levels and memory retention are evaluated, providing valuable data on the cognitive impact of using immersive technologies for cyber situational awareness. 

The findings from these studies consistently indicate that ICSA systems significantly enhance users' performance and cognitive capabilities. By improving threat detection and response times and reducing cognitive load, ICSA systems enable users to manage cybersecurity tasks more effectively. \textcolor{blue} {Table~\ref{tab:ICSAevaluation}} provides a detailed summary of the evaluation mechanisms employed in the existing literature. While we reference all the evaluation mechanisms used, due to space constraints, we have not included every individual study. However, the evaluation methods mentioned are representative of those commonly cited across multiple studies. The table includes the demographics of the users involved, the specific performance and cognition metrics assessed, and the overall results. Consistent findings across studies confirm that immersive technologies improve user performance and cognition.

\begin{landscape}
\begin{table}[t]
\captionof{table}{Evaluation mechanisms for ICSA systems.} 
\label{tab:ICSAevaluation} 
\fontsize{7.5pt}{8pt}\selectfont
     \begin{tabular}{|C{9em}|C{13em}|C{10em}|C{10em}|C{9.5em}|C{15em}|}
         \hline
         
         \textbf{Paper} &  \textbf{Evaluation Mechanism} &  \textbf{Users' Demographics} &  \textbf{Performance Metrics} &  \textbf{Cognition Metrics} &  \textbf{Results}\\ \hline
          
          Mattina et al. \cite{mattina2017marcs} & Questionnaire & & User time-on-task; User fact recall & & Users' performance is improved \\ \hline
          
          Korkiakoski et al. \cite{korkiakoski2021using} & SART Questionnaire; Analysis of Variance with p-value Test & 6 participants; 3 males and 3 females & No. of Completed Tasks & & Understanding of SA depends on gender \\ \hline
          
          Alqahtani et al. \cite{alqahtani2020design} & Questionnaire with five-point Likert Scale & 91 participants; 59\% male and 41\% female; Age between 18 to 65 &  & & AR-based game increases cyber SA \\ \hline

          Beitzel et al. \cite{beitzel2016cognitive} & Capture the flag Exercise; Post-Task Survey; NASA TLX Assessment  & 7 participants; 7 male; Age is between 34 to 60; Ethnicity: Caucasian & Total Elapsed Time; Average Response Time; Countermeasure Failure Rate; Success EOIs; Failed EOIs  & Mental Demand; Physical Demand; Temporal Demand; Frustration & AR improves both performance and cognition of users\\ \hline

          Seo et al. \cite{seo2019using} & Questionnaire & 25 participants & & Memory Test & Interactive immersion in VR is beneficial for long-term memorization of cyber SA \\ \hline
          
          Chu et al. \cite{chu2019data} & Questionnaire; Interview & 6 participants with no cybersecurity knowledge & & Memory Test & Cyber SA training through VR is more engaging than video training.\\ \hline
          
          Jin et al. \cite{jin2018game} & Questionnaire with 5-point Likert Scale; Analysis of Variance with p-value Test  & 181 participants; 123 male and 58 female & & & Immersive game-based learning for cyber SA is more effective for males than females \\ \hline
          
          Kabil et al. \cite{kabil2020alert} & SUS Usability Questionnaire; Analysis of Variance with p-value Test; Cybersickness Questionnaire  & 30 participants with no cybersecurity knowledge &  & Physiological Disorder & Users had good performance with no discomfort \\ \hline
          
          Salazar et al. \cite{salazar2013enhancing} & Questionnaire with 5-point Likert Scale & 208 participants; Age between 14 to 19 & Knowledge Acquisition; Vulnerability Detection; Defense Preparation & Confidence in Technology & AR-based games improve both performance and cognition \\ \hline
          
          Rana et al. \cite{rana2014exploring} & Post-Task Quiz and Survey with Statistical Analysis (t-test) & 100 participants with cybersecurity knowledge & & & VR cybersecurity training is more effective than video-based methods \\ \hline
          
          Delcombel et al. \cite{delcombel2021cybercopter} & SUS Usability Questionnaire & 8 participants; 3 male and 5 female; Age: 23 to 30 & Task Completion Time; Task Accuracy & & Users' task performance is enhanced in immersive environments \\ \hline
          
          Kaleem et al. \cite{Kaleem2019security} & Pre- Post-Task Survey with 5-point Likert Scale and Statistical Analysis & 20 participants & & & AR has a positive impact on cybersecurity learning \\ \hline
          
          Kasurinen et al. \cite{kasurinen2017usability} & Pre- Post-Task Survey & & Task Completion Time; No. of Unforced Errors & & VR learning environments is beneficial for understanding cyber SA \\ \hline
          
          Puttawong et al. \cite{puttawong2017vrfiwall} & Pre- Post-Task Survey & Participants has cybersecurity knowledge & & & VR environments are productive for cybersecurity education \\ \hline
          
          Kommera et al. \cite{kommera2016smart} & Pre- Post-Task Survey; Satisfaction Survey & & & & AR provides insights in cybersecurity forensic education \\ \hline

   \end{tabular}
\end{table}

\end{landscape}



\section{Discussion} \label{discussion}

This section reports an analysis of our research findings described in Sections \ref{visualization&Interactions}, \ref{ICSA_Levels}, and \ref{evaluation}. We develop a reference framework by mapping the ICSA visualization and interaction techniques with different levels of ICSA (perception, comprehension, and projection). This framework aids researchers and practitioners in designing and evaluating ICSA systems using immersive technologies for cyber situational awareness. Furthermore, based on our research findings, we suggest future research directions to further advance the field of ICSA systems.

\subsection{A Reference Framework for Designing and Analyzing ICSA Systems} \label{framwork}

Unlike the traditional cyber SA frameworks \cite{onwubiko2016understanding}, there exists no framework for the designing elements (i.e., interaction and visualization techniques) of ICSA systems to create perception, comprehension, and projection of cyber situational awareness. This gap in the literature and practice presents significant challenges for developers and practitioners working on ICSA systems, as they lack a standardized guide to create systems that effectively enhance the perception, comprehension, and projection of cyber situational awareness. The absence of such a framework leads to several specific issues. Developers, when faced with a multitude of available features and options, often find it difficult to identify the most suitable visualization and interaction techniques for their specific objectives, such as improving user comprehension. This difficulty can result in inefficiencies and increased costs during both the design and operational phases of ICSA systems. Without a clear framework, the process of trial and error becomes more prevalent, which can delay development timelines and inflate budgets. Also, the lack of a structured approach can lead to inconsistencies in system performance and user experience, undermining the efficacy of ICSA systems in real-world applications. For instance, a developer aiming to design an ICSA system with a primary focus on enhancing user perception might struggle to select the appropriate visualization techniques that provide clear and immediate insights into cyber threats. Similarly, interaction techniques that facilitate quick and intuitive user responses might be overlooked or misapplied. These challenges are exacerbated when considering the need to balance multiple design elements simultaneously, such as ensuring that the system is both comprehensive and comprehensible.

\begin{figure}[t]
  \centering
  \includegraphics[width=0.9\columnwidth]{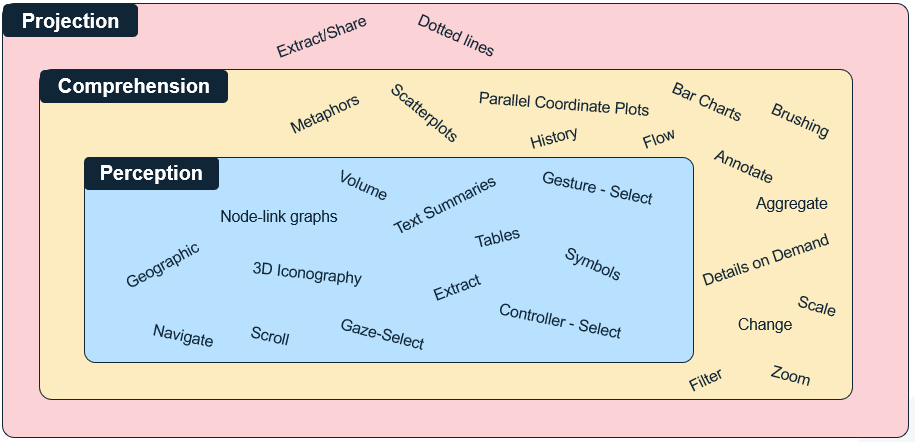}
  \caption{Reference Framework for Designing and Analyzing ICSA Systems}
  \label{fig:frameworkk} 
  \vspace{-15pt}
\end{figure}

To address these significant challenges, we have undertaken the task of mapping visualization and interaction techniques (detailed in Section \ref{visualization&Interactions}) to the different levels of ICSA (outlined in Section \ref{ICSA_Levels}). This mapping process has culminated in the development of a reference framework that categorizes existing interaction and visualization techniques according to the three critical levels of ICSA: perception, comprehension, and projection. \textcolor{blue}{Fig. \ref{fig:frameworkk}} illustrates this framework, demonstrating the immersive visualization and interaction techniques applicable at each level of ICSA. For instance, basic visualization techniques such as \textit{volume} and interaction features like \textit{select} are essential components at the perception level of cyber situational awareness. These techniques are utilized within our framework to enhance the perception level of ICSA, providing users with clear and immediate insights into cyber threats. Our reference framework is designed to aid developers and practitioners in identifying the most suitable visualization and interaction techniques when designing and operating ICSA systems for specific purposes, whether it be for perception, comprehension, or projection. By offering a structured approach, this framework ensures that the selection and implementation of these techniques are efficient and effective, ultimately enhancing the overall design and functionality of ICSA systems. Moreover, the framework plays a crucial role in facilitating anomaly detection and mitigation processes within ICSA systems. Categorizing immersive techniques based on their applicability to different levels of situational awareness enables the creation of more robust systems that can promptly identify and respond to anomalies. This capability is particularly important in the dynamic field of cybersecurity, where timely and accurate situational awareness is critical to preventing and addressing cyber threats.

\subsection{Future Research Areas}

Through the analysis of our research findings, we have identified potential future research directions for researchers aimed at improving the existing state of ICSA systems.

\textbf{Projection of ICSA.} Prediction of cyber situations is an integral component of situational awareness, enabling users to forecast imminent situations in cyberspace and formulate optimal response plans. However, as highlighted by our proposed framework (\textcolor{blue}{Fig. \ref{fig:frameworkk}}), there is a significant lack of immersive visualization and interaction techniques tailored for the projection phase of ICSA. Therefore, we encourage researchers and practitioners of ICSA systems to focus on developing advanced techniques for the projection phase. This focus will help users obtain a holistic view of cyber situational awareness within immersive environments, thereby enhancing their ability to anticipate and respond to cyber threats effectively.

\textbf{Integrating Advanced Immersive Visualization and Interaction Techniques.} Immersive technologies offer a wide array of advanced visualization and interaction techniques that have yet to be fully integrated into the realm of ICSA. These innovative techniques include, but are not limited to, flow visualizations, Kohonen map representations, heatmaps, import interactions, and drive interactions.

Flow visualizations \cite{homps2020revivd} are particularly valuable for illustrating the movement and interaction of data within a network, helping users to quickly identify unusual patterns and potential threats. Kohonen map representations \cite{kraus2022immersive}, also known as self-organizing maps, provide a means of visualizing high-dimensional data in a two-dimensional space, which can be instrumental in identifying clusters and anomalies in large datasets. Heatmap visualizations \cite{kraus2020assessing} employ a three-dimensional approach to represent data, where color gradients indicate data intensity and the height of the map reflects the magnitude of specific metrics. This method offers an intuitive understanding of data variations and hotspots. Import interactions \cite{benko2004collaborative} involve the ability to seamlessly bring external data into the immersive environment, enhancing the user's capacity to analyze and correlate diverse data sources. Drive interactions \cite{fonnet2019survey} refer to the techniques that enable users to navigate through the virtual space effectively, providing them with a more immersive and interactive experience.

The incorporation of these advanced visualization and interaction techniques into the ICSA domain can significantly enhance the capability of developers and practitioners to design and operate ICSA systems. By leveraging these techniques, users can gain deeper insights into cyber threats and network activities, enabling more effective monitoring, analysis, and response. Furthermore, we strongly advocate for the continuous development and integration of new visualization and interaction techniques tailored specifically for cyber situational awareness in immersive environments. This ongoing innovation is essential to keep pace with the evolving nature of cyber threats and to ensure that ICSA systems remain robust, intuitive, and effective.

\textbf{Large-scale User Study:} As described in Section \ref{evaluation}, most ICSA systems are evaluated through user studies involving a relatively small number of participants. This approach raises concerns regarding the validity and generalizability of these systems. For example, an ICSA system that has been evaluated by only a handful of users may not be applicable or effective in large-scale infrastructures with numerous operators and diverse user requirements. The limitations of small-scale studies include a narrow scope of user feedback, which might not capture the full range of potential issues and strengths of the system. This can lead to design choices that work well in controlled, limited settings but fail to perform in more complex, real-world environments. The feedback from a small group may not reflect the varied experiences and needs of a broader user base, which can result in a lack of scalability and adaptability in ICSA systems. 

To address these concerns, we propose conducting large-scale user studies for analyzing and testing the design and development of ICSA systems. Large-scale studies involve a significant number of participants, ideally representative of the diverse user base that the system aims to serve. This includes users with different levels of expertise, from various demographic backgrounds, and operating in different environments. Conducting large-scale user studies offers several key benefits. Enhanced validity is achieved with a larger and more diverse participant pool, resulting in findings that are more likely to be valid and reliable, accurately reflecting the needs and behaviors of a broad user base. Comprehensive feedback from a larger number of participants provides extensive and varied insights, identifying potential issues and areas for improvement that might not be evident in small-scale studies. This feedback is crucial for refining the system to ensure it meets user needs effectively. Improved generalizability is another benefit, as results from large-scale studies are more likely to apply to different contexts and user groups, enabling confident deployment of ICSA systems in various settings, from small organizations to large macro infrastructures. Finally, robust design and development are supported by large-scale user studies, which uncover insights that inform better practices, leading to the creation of more robust, scalable, and user-friendly ICSA systems capable of performing well under diverse conditions and in complex environments.

\subsection{Industrial Implications of ICSA Systems.}

Our research paper highlights the transformative potential of immersive technologies within the cybersecurity industry, drawing parallels with their successful adoption in gaming, entertainment, education, and healthcare. Our findings indicate that immersive technologies, such as augmented reality and virtual reality, are increasingly being utilized for cybersecurity awareness and technical training, offering engaging and effective learning experiences. These technologies create realistic simulation environments that provide cybersecurity professionals with hands-on experience, thereby enhancing their skills and preparedness for real-world threats. Moreover, our research shows that immersive technologies facilitate the visualization of complex network traffic, security alerts, cyber threats, and network architectures, making it easier for security teams to comprehend and manage intricate cybersecurity landscapes. They also enhance Security Operations Center and Network Operations Center capabilities by enabling real-time collaboration and data visualization, allowing teams to respond swiftly and efficiently to security incidents. Furthermore, our paper highlights how these technologies aid in network and security troubleshooting by providing an interactive platform for real-time collaboration and detailed data analysis.

While our research underscores the numerous benefits of adopting immersive technologies in the cybersecurity industry, we also emphasize the need for practitioners to be aware of the potential security and privacy risks. The use of AR and VR tools can introduce new vulnerabilities that attackers might exploit \cite{alismail2022systematic}. For instance, these tools themselves can have security weaknesses, posing risks to valuable data collected from individuals interacting with these tools, such as behavioral and biometric data. If not properly secured, this data can become a target for cybercriminals, leading to breaches of privacy. Therefore, we stress the importance of implementing stringent privacy and security measures to mitigate these risks. This includes data encryption to protect sensitive information, robust privacy policies to ensure responsible data handling, and multifactor authentication to enhance security access controls. Endpoint security measures are also critical in safeguarding devices used in immersive environments. Furthermore, maintaining proper security configurations is essential to secure all immersive technology tools against potential exploits. By balancing the innovative advantages of immersive technologies with rigorous security and privacy protocols, organizations can leverage these tools to enhance their cybersecurity capabilities while minimizing associated risks, as highlighted in our comprehensive analysis.



\section{Conclusion} \label{conclusion}

Immersive technologies are increasingly used to create cyber SA through interactive visualizations for cybersecurity analysis, education, and training. Therefore, we have collected, investigated, and synthesized the body of knowledge on ICSA systems. In this survey, we have described 11 visualization techniques, 9 interaction features, three levels of cyber SA, and evaluation mechanisms for ICSA systems. Moreover, we have critically analyzed the research findings, which enables us to: (i) propose a reference framework for designing and analyzing ICSA systems by mapping immersive visualization and interaction techniques to the different levels of ICSA; (ii) propose future research directions for advancing the state-of-the-art of ICSA systems; and (iii) highlight the industrial implications of ICSA systems.

This survey facilitates researchers and practitioners of ICSA systems in many ways. For researchers, we have identified several future research areas for advancing the state of the art on ICSA systems. For example, the projection phase of ICSA needs innovative visualization and interaction techniques that help users in forecasting imminent cyber situations. Similarly, our survey highlights the need for large-scale user studies for providing tested and validated ICSA systems to end-users. For practitioners, this survey presents a framework that categorizes immersive visualization and interaction techniques according to the levels of ICSA. This facilitates practitioners in selecting suitable visualization and interaction techniques for designing and operating ICSA systems for specific purposes (e.g., perception). We hope this survey will provide researchers and practitioners with innovative ways and inspirations to use immersive technologies for cyber SA.


\bibliographystyle{ACM-Reference-Format}
\bibliography{acmart}

\end{document}